\documentstyle[11pt]{article}
\begin{document}

\bigskip
\begin{center}
{\Large {\bf Late-time Entropy Production from Scalar Decay and
Neutrino Decoupling}}
\newline

{\sf Paramita Adhya $^{a,\!\!}$
\footnote{E-mail address: rinkyadh@cal2.vsnl.net.in}} and
{\sf D. Rai Chaudhuri $^{a,\!\!}$
\footnote{E-mail address: allraich@vsnl.net}}

\smallskip
$^a${\it Department of Physics, Presidency College, 86/1, College
Street,\\ Calcutta 700073, India}\\
\medskip
{\bf Abstract}
\end{center}
{\small Late-time entropy production from scalar decay arises in
scenarios like thermal inflation, proposed to dilute long-lived,
massive fields like the gravitino and the moduli. The scalar decay
may continue into Mev-scale temperatures and affect BB
nucleosynthesis. The effect of such entropy production on electron
neutrino decoupling is studied. A lower bound of about $10^{-22}$ Gev
is estimated for the scalar decay constant, such that, for higher
values of the decay constant, standard electron neutrino decoupling
is unaffected.}

\medskip

\begin{center}
PACS Numbers: 98.80.Cq, 98.70.Vc\\ Keywords: thermal inflation,
neutrino decoupling, scalar decay, entropy production

\end{center}

\newpage
\section{\bf  Introduction}Supersymmetric and string theory models
throw up long-lived, massive fields like the gravitino, the Polonyi,
the moduli and the dilaton \cite{giu}. Stability of the corresponding
particle may lead to over-abundance, while decay during baryogenesis
or BBN nucleosynthesis may disturb $\eta$ or nuclear abundances too
much \cite{intro}. Thermal inflation \cite{flaton} has been proposed
as a method of diluting away troublesome fields. A scalar field, the
flaton, is used to start inflation at a temperature of about $10^7$
Gev. The inflation stops at a temperature of the order of the flaton
mass, typically $\le$ $10^3$ Gev. Such a scalar field goes on
decaying into Mev-scale temperatures with potential trouble for BBN
nucleosynthesis.
\par The effect of flaton decay on the effective number of neutrino
species, $N_{eff}$, and the neutrino distribution function during
nucleosynthesis has been recently studied in detail
\cite{kawa,kawa1}. Here, we study the effect of scalar decay on
decoupling of the usual left-handed, massless neutrinos. As the
phenomenological details of such a decaying particle are still quite
open, we consider the general effect of late time entropy production
on electron neutrino decoupling, assuming that the decaying scalar
once dominated the energy density of the universe. The electron
neutrino decoupling temperature is a sensitive input of BBN, and we
estimate lower bounds on the scalar decay constant, demanding that
this decoupling temperature should not be disturbed.

\section {\bf Entropy Generation and Decoupling}  The entropy
generation rate is calculated from \[
dS_{\Phi}=-\frac{d(a^3\rho_{\phi})}{T}, \] where $\rho_{\phi}$ is the
scalar energy density at radiation temperature T, all massless
particles being collectively called radiation. Let $\Gamma $ be the
scalar decay constant. Writing $\Phi=a^3\rho_{\phi},$ the equation
for the evolution of the scalar energy density is
$\dot{\Phi}=-\Gamma\Phi,$ with the solution $ \Phi=\Phi_E e^{-\Gamma
(t-t_E)}\label{eq:Phi}\cite{scherrer},$ where $\Phi_E$ is the value
of $\Phi$ at a fiducial time $t=t_E .$ This gives
\begin{equation}\dot{S_{\Phi}}=(\Gamma /T)\Phi_E e^{-\Gamma
(t-t_E)}\label{eq:Sdot}.\end{equation}
\par What is the standard situation in the absence of scalar decay?
In the era of neutrino decoupling, the electron neutrinos interact
via charged and neutral currents, while muon and taon neutrinos
interact via neutral currents alone. So, the muon and taon neutrinos
decouple at a slightly higher temperature than the electron
neutrinos. Then, the change in electron neutrino ($\nu$ ) number
occurs due to the process $\nu +\bar{\nu}\to e^-+e^+$.\par The
decoupling of electron neutrinos is governed by the integrated
Boltzmann equation \cite{lee,early}\begin{equation}\dot{n} +
3Hn=-<\sigma|v|>(n^2-n_{EQ}^2)\label{eq:leew},\end{equation}where $n$
is the number density of the electron neutrinos, $n_{EQ}$ their
equilibrium number density, and $<\sigma|v|>$ the thermally averaged
cross-section times relative velocity. Assuming an absence of
neutrino degeneracy, we use the Boltzmann distribution function to
give \begin{equation}n_{EQ}=T^3/\pi^2.\label{eq:neq}\end{equation}
Following the usual method of calculation \cite{early,pa1,pa2},
neglecting the masses of the neutrino and the electron,
\begin{eqnarray}<\sigma|v|>=
\frac{8}{\pi}G_F^2[(C_{Ve}+1)^2+(C_{Ae}+1)^2]T^2\nonumber \\ =
\frac{4.112\times 10^{-10}}{Gev^4}T^2.\label{eq:sigmav}\end{eqnarray}
The decoupling may be taken to start when the following relation just
holds.
\begin{equation}-<\sigma|v|>(n^2-n_{EQ}^2)<3Hn\label{eq:cri}.
\end{equation}
\par  Denoting $(a^3)$ times entropy density as $S$, the net
contribution, into the $\nu,\bar{\nu}$ sector, of the covariant
divergence $(1/a^3)\dot{S}$ of the usual entropy density current, due
to the process $e^-+e^+\to \nu +\bar{\nu}$, is
$-2\alpha<\sigma|v|>(n^2-n_{EQ}^2) \cite{bern},$
where$^{}$\footnote{In \cite{bern}, the distribution function of the
decoupling particle is taken as $f(p)=e^{-\alpha(t)
-\beta(t)E(p)}(1+\xi(p))$. For sufficiently strong elastic
collisions, it is shown that $\xi\ll 1$, and the process is well
represented by the time-dependent parameter $\alpha(t)$. Despite the
resemblance, $\alpha$ is not a chemical potential. It parametrises
the distribution and is the {\it same} for particle and
anti-particle, instead of being different in sign.\par The $2$ factor
in $2\alpha$ arises because our equation (\ref{eq:leew}) considers
the change in particle density alone, while $\dot{S}$ considers net
entropy transferred to both particle and antiparticle. \cite{bern}
does not have the $2$ factor as particle and anti-particle are there
considered together as a single species.}
\begin{equation}n=n_{EQ}e^{-\alpha}\label{eq:alpha}.\end{equation} If
there is additional entropy generation due to scalar decay, there
will be a corresponding extra term on the RHS of (\ref{eq:leew}). To
find a lower bound to $\Gamma$, such that scalar decay does not
affect neutrino decoupling, we require this additional term to be
small compared to the term $-<\sigma|v|>(n^2-n_{EQ}^2)$ on the RHS,
at decoupling. To find the lower bound on $\Gamma$, we consider the
worst case scenario when the entire additional entropy due to scalar
decay is transferred$^{}$\footnote{In current models, it is usually
assumed that the scalar cannot decay directly to neutrinos
\cite{kawa}. Any massive decay products of the scalar quickly
decouple and the entropy due to scalar decay passes to the
$e^-,e^+,\gamma$ sector, and thence to the neutrinos via the
annihilation process $e^-+e^+\to \nu +\bar{\nu}$. As electromagnetic
interactions keep the $e^-,e^+,\gamma$ sector very near equilibrium,
almost all the entropy may be supposed to pass to the neutrinos.} to
the neutrinos. Then, if decoupling is to be kept undisturbed, we can
require that the contribution, to the $\nu,\bar{\nu}$ sector, of the
covariant divergence of the entropy current due to scalar decay
should be less than that due to the usual annihilation process, or\[
\frac{1}{a^3}\dot{S_{\Phi}}<-2\alpha<\sigma|v|>(n^2-n_{EQ}^2),\]when
(\ref{eq:cri}) holds. This means that the required lower bound to
$\Gamma$ is to be obtained from the
criterion\begin{equation}\frac{1}{a^3}\dot{S_{\Phi}}<2\alpha(3Hn)
\label{eq:criphi},\end{equation}where
$\alpha $ corresponds$^{}$\footnote{We assume that the electron
neutrino distribution function continues to be parametrised as
$f(p)=e^{-\alpha(t)-\beta(t)E(p)}$ even when there is scalar decay.
It is shown in meticulous detail in ref.\cite{kawa} that, for low
values of the  scalar decay constant, the distribution is not
thermalised. There is a deficit from the thermal distribution. So,
our assumption is tantamount to an approximation of the deficit by a
uniform factor $e^{-\alpha}$, with $\alpha>0$.}  to its decoupling
value, in the absence of scalar decay,  worked out from the equality
corresponding to (\ref{eq:cri}).

\section{\bf  Estimating the Entropy Generation} (\ref{eq:Sdot})
indicates that $\dot{S_{\Phi}}$ depends on $\Phi_E$. In the absence
of definite phenomenological values, $\Phi_E$ is to be estimated
indirectly. We are interested in the era when radiation domination
has begun, and (\ref{eq:Sdot}) is dominated by the exponential. Then,
the pre-exponential may be estimated approximately, to an order of
magnitude. So, in the pre-exponential, we define the initial epoch
$t_E$ by putting $\rho_{\phi E}=\rho_{RE} $, $\rho_R $ being the
radiation energy density. Radiation domination is taken to start
after this epoch. Further, we neglect entropy generation while
estimating the pre-exponential in (\ref{eq:Sdot}), and put
\begin{eqnarray}\Phi_E=a_E^3\rho_{\phi E}=a_E^3\rho_{R
E}=(\pi^2/30)g_E^*a_E^3T_E^4\nonumber\\ =
(\pi^2/30)g^*a^3T^3T_E\label{eq:PhiE},\end{eqnarray}where $g^*,a,T$
refer to the epoch of decoupling.  \subsection {\it Estimate of $T_E$
} We assume that when  the neutrinos decouple, $\rho_{\phi}\ll\rho_R
$, such that $\rho_{\phi}/\rho_R$ cannot be neglected, but its higher
powers can. It is possible to show \cite{pa}, in fact, that
decoupling is not possible when $\rho_{\phi}\gg\rho_R.$ In the era
$\rho_{\phi}\ll\rho_R $ of incomplete radiation domination, the
Friedmann equation is put in the form
\begin{equation}H^2=\frac{8\pi R}{3M_{Pl}^2a^4}(1+\frac{\Phi
a}{R})\label{eq:FriR},\end{equation}where $R=a^4\rho_R.$ 
If, well into this epoch, the correction term $\Phi a/R$ on the RHS
of (\ref{eq:FriR}) is neglected, the full radiation domination
relations are found:\begin{eqnarray}H &=&\frac{1}{2t},\mbox{ and
}\nonumber \\ a&=&At^{\frac{1}{2}},\label{eq:rdom}\end{eqnarray}$A$
being a constant.\par The $\Phi$ evolution equation  has, as
solution, a falling exponential in $t$, viz. $\Phi\sim e^{-\Gamma
t}$. Instead of taking the falling exponential in $t$ directly, a
suitable approximation to the correction term $\Phi a/R$ on the RHS
of (\ref{eq:FriR}) is first worked out. Let $t_0$ be a sufficiently
late epoch, when $\Phi=\Phi_0\approx 0$.  Then, for use only in the
correction term $\Phi a/R$, one takes
\begin{eqnarray} \Phi - \Phi_0=
\tilde{\Phi}(\frac{1}{t})-\tilde{\Phi}(\frac{1}{t_0})=
\frac{d\tilde{\Phi}}{d\frac{1}{t}}|_{t_0}(\frac{1}{t}-\frac{1}{t_0}).
\nonumber\end{eqnarray}
Neglecting $\Phi_0,1/t_0$ compared to $\Phi,1/t$, respectively, an
approximation\begin{equation}
\Phi\approx\frac{B}{t},\label{eq:Phiapprox}\end{equation}will be used
only in the correction term $\Phi a/R$, i.e., 
in the correction term, the falling exponential will be approximated
by a rectangular hyperbola. B is a constant. \par A similar
approximation is considered for $R$.  From\[ \frac{\partial}{\partial
t}[a^3(\rho_{\phi}+\rho_R)]+p_R\frac{\partial}{\partial t}a^3=0,\]
one obtains\begin{equation}
\dot{R}=a\Gamma\Phi\label{eq:Rdot}.\end{equation} It ought
to be mentioned that $R$ refers to the total radiation present, and
not only to that produced by decay. However, the change in $R$ is due
to $\phi$ decay and consequent entropy production. In the absence of
this decay, $\dot{R}=0$. Using (\ref{eq:rdom}) and
(\ref{eq:Phiapprox}) in (\ref{eq:Rdot}), and, integrating, one
obtains approximately, for use only in the correction term $\Phi
a/R$,\[R-R_E\approx 2AB\Gamma(t^{\frac{1}{2}}-t_E^{\frac{1}{2}}).\]If
$t_E$ is sufficiently early compared to $t$, and there is
sufficiently copious radiation production since $t_E$, it is
sufficient to take \[R\approx 2AB\Gamma t^{\frac{1}{2}}\] in the
correction term $\Phi a/R$. (\ref{eq:rdom}) and (\ref{eq:Phiapprox})
are now used to give, in the correction term,
\begin{eqnarray}
x &=&\frac{\Gamma}{H}\nonumber \\ & \approx & \frac{R}{\Phi
a}.\label{eq:defx}\end{eqnarray}\par Introducing the variable $x$ in
(\ref{eq:FriR}), we get (\ref{eq:FriR}) in the form
\begin{eqnarray}H=\frac{\Gamma}{x}\nonumber \\ 
=\frac{4.461\times 10^{-19}}{Gev}T^2(1+\frac{1}{x})^{\frac{1}{2}}
\label{eq:Hx},\end{eqnarray} putting $g^*=10.75$. 
(\ref{eq:Hx}) is a good equation for $x\gg 1.$ But, we approximate
$T_E$ in the pre-exponential term $\Phi_E$ of (\ref{eq:Sdot}), by
putting $x=x_E=1$ in (\ref{eq:Hx}) (corresponding to
$\rho_{\phi}=\rho_R$, or $ \Phi a = R$, at $ t=t_E$) to get
\begin{equation}T_E=\frac{\sqrt{\Gamma}}{\sqrt{\sqrt{2}\times
4.461\times 10^{-19}/Gev}}\label{eq:TE}.\end{equation}

\section{\bf Numerical Results and Conclusions}
Now, BB nucleosynthesis is consistent with a neutrino decoupling
temperature $T_D \sim $ a few Mev.  Here, we take $T_D=1,2,3$ Mev,
and find corresponding values of $\alpha$, in the absence of scalar
decay, from the equality corresponding to (\ref{eq:cri}), using
(\ref{eq:neq}), (\ref{eq:alpha}) and $H=4.461\times 10^{-19}T^2$, the
last being the usual radiation domination value of $H$. The results
are shown in the first two columns of Table
I.\begin{table}\bf\begin{center}\begin{tabular}{|l|c|c|r|}
\hline $T_D/Mev$ & $\alpha $ & $x_D$ & $\Gamma/(10^{-24}Gev)$ \\
\hline 1 & 3.4705 & 17.19 & 7.9\\
\hline  2 & 1.447 & 14.36 & 26.5\\
\hline 3 & 0.5644 & 14.505 & 60.2\\ \hline \end{tabular}\\ 
{\small \sf Table I: Neutrino $\alpha $ values at decoupling for
different decoupling temperatures\\ and the lower bound on the decay
constant $\Gamma$}\end{center}\end{table}
\par Next, we consider scalar decay. The value of $T_E$ from
(\ref{eq:TE}) is used in (\ref{eq:PhiE}) and the latter is put into
(\ref{eq:Sdot}). Utilising (\ref{eq:Hx}) to relate the temperature
and $x$ where necessary, and estimating $t_E$ from $H=1/(2t),
(\ref{eq:defx}), \, \mbox{and} \, x_E=1$, the equality corresponding
to the relation (\ref{eq:criphi}) is now solved for $x=x_D$, keeping
the same values of $\alpha$ already found to correspond to decoupling
temperatures $T_D=1,2,3$ Mev in the absence of scalar decay. The
results are in the third column of Table I. Finally, $\Gamma$ is
found corresponding to these values of $x_D$ from (\ref{eq:Hx}) for
$T_D=1,2,3$ Mev. The form of $\dot{S_{\Phi}}$ and the inequality
(\ref{eq:criphi}) show that these values correspond to a lower bound
on $\Gamma$. \par We conclude that if the entropy produced by scalar
decay is not to disturb the neutrino decoupling temperature, the
lower bound on the scalar decay constant must be of the order of
$8\times 10^{-24}Gev, \,26.5\times 10^{-24}Gev, \, 60\times
10^{-24}Gev$, corresponding to neutrino decoupling temperatures of
$1,2,3$ Mev. So, neutrino decoupling is unaffected, and BB
nucleosynthesis safe on this count, if the scalar decay constant is
greater than about $10^{-22}$ Gev, corresponding to a reheating
temperature of 8.6 Mev. \medskip \\ {\bf Acknowledgements} \par PA
wishes to thank the authorities of Presidency College, Kolkata, for
facilities. DRC wishes to thank Rahul Biswas for help in locating
reference material.

\end{document}